\documentclass[fleqn,twoside]{article}
\usepackage{espcrc2}

\usepackage{graphicx}
\usepackage[figuresright]{rotating}

\newcommand{\AmS}{{\protect\the\textfont2
  A\kern-.1667em\lower.5ex\hbox{M}\kern-.125emS}}
  
\title{Ultrahigh Energy Cosmic Rays and Neutrinos}

\author{A. V. Olinto\address[UC]{Department of Astronomy and Astrophysics, 
Kavli Institute for Cosmological Physics, \\
The University of Chicago, Chicago, IL 60637, USA}%
        \thanks{This work was supported by the NSF grant PHY-0758017,  the KICP through grant NSF PHY-0551142, and an endowment from the Kavli Foundation in the U.S., and the CNRS in France.}
        K. Kotera\addressmark[UC]
        and 
        D. Allard\address{Laboratoire Astroparticules et Cosmologie  (APC), Universit\'e Paris 7/CNRS, 10 rue A. Domon et L. Duquet, 75205 Paris Cedex 13, France.}}
       
\begin{document}

\begin{abstract}
The observation of neutrinos from cosmic accelerators will be revolutionary. High energy neutrinos are closely connected to ultrahigh energy cosmic rays and their sources. Cosmic ray sources are likely to produce neutrinos and the propagation of ultrahigh cosmic rays from distant sources can generate PeV to ZeV neutrinos.  We briefly review recent progress on the observations of ultrahigh energy cosmic rays and their implications for the future detections of high energy neutrinos.
\vspace{1pc}
\end{abstract}

% typeset front matter (including abstract)
\maketitle

\section{Introduction}

The detection of neutrinos from astrophysical sources will open a new window onto the workings of the highest energy cosmic accelerators and provide new ways to test neutrino interactions. Neutrino fluxes can be estimated if sources of ultrahigh energy cosmic rays (UHECRs) are well understood. However, the sources of UHECRs, their location, cosmological evolution, and maximum energy, as well as the injected composition, are presently unknown.

Neutrinos can be generated at the sources of cosmic rays or during the propagation of UHECRs as they interact with the ambient photon backgrounds \cite{BZ69,Stecker79}. The neutrinos generated during UHECR propagation, named cosmogenic neutrinos, represent an almost guaranteed flux and  have encouraged efforts to detect them for decades (see, e.g.,  \cite{AM09}). One important assumption for the existence of cosmogenic neutrinos, that cosmic rays are extragalactic at the highest energies, has been verified by the detection of a feature consistent with the Greisen-Zatsepin-Kuzmin (GZK) cutoff \cite{G66,ZK66} in the cosmic ray spectrum \cite{Abbasi09,Abraham:2008ru} 
and by the indication of anisotropies in the cosmic ray sky distribution at the highest energies  \cite{Auger1,Auger2}. 
These findings herald a possible resolution to the mystery behind the origin of UHECRs and the possibility of detecting ultrahigh energy neutrinos in the near future. 

This optimistic view has been recently damped by the indication that UHECRs may be dominated by heavier nuclei \cite{Abraham:2010yv}. The cosmogenic neutrino flux expected from heavy cosmic ray primaries can be much lower than  if the primaries are protons at ultrahigh energies, making a detection extremely challenging for current observatories.  Conversely, if neutrinos are observed, they will test specific sets of cosmic ray source parameters. 

Below we review the current state of UHECR observations. In section 3, we describe the impact of these observations on predicted neutrino fluxes and conclude in section 4.

\section{Recent process in UHECRs}

The dominant component of cosmic rays observed on Earth are believed to originate in Galactic cosmic accelerators. A transition from Galactic to extragalactic cosmic rays should occur somewhere between  a PeV ($\equiv 10^{15}$ eV) and a few EeV ($\equiv 10^{18}$ eV). Above a few EeV,  the so-called ultrahigh energy cosmic rays are most likely extragalactic. These are observed to reach energies that exceed $10^{20}$ eV posing some interesting and challenging questions: Where do they come from? How can they be accelerated to such high energies? What kind of particles are they? What is the spatial distribution of their sources? What do they tell us about these extreme cosmic accelerators? How strong are the magnetic fields that they traverse on their way to Earth? How do they interact with the cosmic background radiation? What secondary particles are produced from these interactions? What can we learn about particle interactions at these otherwise inaccessible energies? Reviews on the science of ultrahigh energy cosmic rays can be found in \cite{KKAO11,Letessier11}, while for high energy neutrinos see \cite{AM09}.
 
\begin{figure}[!t]
\centerline{\includegraphics[height=0.3\textheight]{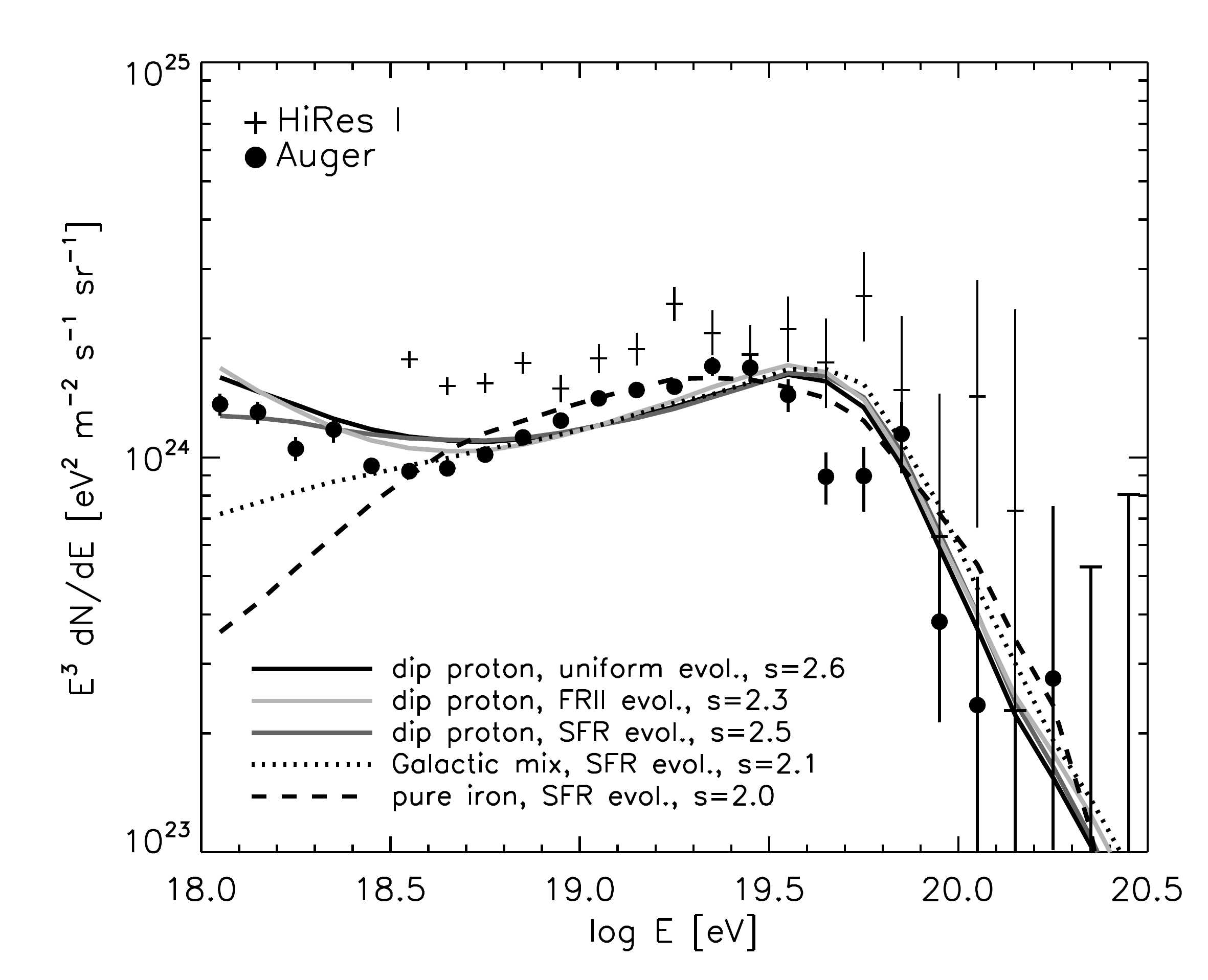}}
\caption{Spectrum of UHECRs multiplied by $E^3$ observed by HiRes I \cite{Abbasi09} and Auger \cite{Abraham10}. The displayed error bars are statistical errors while the reported  systematic error on the absolute energy scale is about 22\%. Overlaid are simulated spectra obtained for different models of the Galactic to extragalactic transition and different injected chemical compositions and spectral indices, $s$. }
\label{uhecrSpec}
\end{figure}

Recent observations of UHECRs reveal a spectrum whose shape supports the long-held notion that sources of UHECRs are extragalactic. As shown in Figure  \ref{uhecrSpec}, the crucial spectral feature recently established at the highest energies is a steep decline in flux above about 30 EeV. This feature was first established by the HiRes observatory \cite{Abbasi09} and confirmed with higher statistics by the Pierre Auger Observatory \cite{Abraham:2008ru}. The decline in flux is reminiscent of the effect of interactions between extragalactic cosmic rays and the cosmic background radiation, named the Greisen-Zatsepin-Kuzmin (GZK) cutoff \cite{G66,ZK66}, which causes cosmic ray protons above many tens of EeV to lose energy via pion photoproduction off cosmic backgrounds while cosmic ray nuclei photodissociate. This feature was not seen in earlier observations with the AGASA array \cite{Takeda98}. Another important feature shown in Figure \ref{uhecrSpec}  is the hardening of the spectrum at a few EeV, called the {\it ankle}, which may be caused by the transition from Galactic to extragalactic cosmic rays or by propagation losses if UHECRs are mostly protons.

Figure \ref{uhecrSpec} also shows the observed spectrum fit by different models of UHECR sources (adapted from \cite{KKAO11}). In the mixed composition and iron dominated models   \cite{Allard07}, the ankle indicates a transition from Galactic to extragalactic cosmic rays, the source evolution is similar to the star formation rate (SFR),  and the injection spectra are relatively hard ($s\sim 2 -2.1$). In the proton dominated models in the figure, the ankle is due to pair production propagation losses \cite{BG88}, named ``dip transition models" \cite{BGG06}, and the injection spectra are softer for a wide range of evolution models. Models with proton primaries can also fit the spectrum  with harder injection with a transition from Galactic to extragalactic at the ankle.

The confirmed presence of a spectral feature similar to the predicted GZK cutoff, settles the question of whether acceleration in extragalactic  sources can explain the high-energy spectrum, ending the need for exotic alternatives designed to avoid the GZK feature. However, the possibility that the observed softening of the spectrum is mainly due to the maximum energy of acceleration at the source, $E_{\rm max}$,  is not as easily dismissed. A confirmation that the observed softening {\it is} the GZK feature,  awaits supporting evidence from the spectral shape, anisotropies, and composition and the observation of produced secondaries such as neutrinos and photons.

The landmark measurement of a flux suppression at the highest energies encourages the search for sources in the nearby extragalactic universe using the arrival directions of trans-GZK cosmic rays (with energy above $\sim$ 60 EeV). Above GZK energies, observable sources must lie within about 100 Mpc, the so-called GZK horizon or GZK sphere. At  trans-GZK energies, light composite nuclei are promptly dissociated by cosmic background photons, while protons and iron nuclei  may reach us from sources at distances up to about 100 Mpc. Since matter is known to be distributed inhomogeneously within this distance scale, the cosmic ray arrival directions should exhibit an anisotropic distribution above the GZK energy threshold, provided intervening magnetic fields are not too strong. At the highest energies, the isotropic diffuse flux from sources far beyond this GZK horizon should be strongly suppressed.

The most recent discussion of anisotropies in the sky distribution of UHECRs began with the report that the arrival directions of the 27 cosmic rays observed by Auger with energies above 57 EeV exhibited a statistically significant correlation with the anisotropically distributed galaxies in the 12th VCV \cite{VC06} catalog of active galactic nuclei (AGN)  \cite{Auger1,Auger2}. The correlation was most significant for AGN with redshifts $z <$ 0.018 (distances $< $ 75 Mpc)  and within 3.1$^\circ$ separation angles. An independent dataset confirmed the anisotropy at a confidence level of over 99\% \cite{Auger1,Auger2}. The prescription established by the Auger collaboration tested the departure from isotropy given the VCV AGN coverage of the sky, not the hypothesis that the VCV AGN were the actual UHECR sources. A recent update of the anisotropy tests with 69 events above 55 EeV \cite{Abreu10} shows that the correlation with the VCV catalog is not as strong for the same parameters as the original period. The data after the prescription period shows a departure from isotropy at the 3$\sigma$ level. With the currently estimated correlation fraction of 38\%, a 5$\sigma$ significance will require at least another four years of Auger observations \cite{Abreu10}. No corresponding correlation was observed in the northern hemisphere by HiRes \cite{Abbasi08corr} where the distribution of their 13 trans-GZK events is consistent with isotropy.

The anisotropy reported by the test with the VCV catalog  may indicate the effect of the large scale structure in the distribution of source harboring galaxies or it may be due to a nearby source. An interesting possibility is the cluster of Auger events around the direction of Centaurus A, the closest AGN (at $\sim$  3.8 Mpc). Only much higher statistics will tell if Cen A is the first UHECR source to be identified. 

The third key measurement that can help resolve the mystery behind the origin of UHECRs is their composition as a function of energy observed on Earth. Composition measurements can be made directly up to energies of $\sim$ 100 TeV with space-based experiments. For higher energies, composition is derived from the observed development and particle content of the extensive airshower created  by the primary cosmic ray when it interacts with the atmosphere. 

Observations of shower properties show the dominance of light nuclei around a few EeV. A surprising trend occurs in data by the Auger Observatory above 10 EeV, a change toward heavy primaries is seen both in average position of the maximum of the showers  as well as in the rms fluctuations about the mean up to 40 EeV. As a mixture of different nuclei would increase the rms fluctuations, the observed narrow distribution argues for a change toward a composition dominated by heavy nuclei. In contrast, the HiRes measurement of fluctuations  remains closer to light primaries up to around 50 EeV. The two results are consistent within quoted errors, so the situation is currently unclear. 

Changes to hadronic interactions from current extrapolations provide a plausible alternative interpretation to the observed shower development behavior. Auger probes interactions above 100 TeV center of mass, while hadronic interactions are only known around a TeV. The observation of anisotropies and secondary particles (neutrinos and gamma-rays) can lead to astrophysical constraints on the composition of UHECRs, opening the possibility for the study of hadronic interaction cross sections, multiplicities, and other interaction parameters at hundreds of TeV.

The detailed composition of UHECRs is still to be understood, but it is clear that primaries are not dominated by photons \cite{Aglietta:2007,Abraham:2009qb} or neutrinos \cite{Auger_nu09,Abbasi08neu}. Limits on the photon fraction place stringent limits on models where UHECRs are generated by the decay of super heavy dark matter and topological defects. Unfortunately, the uncertainties on the UHECR source composition, spectrum, and redshift evolution translates to many orders of magnitude uncertainty in the expected cosmogenic neutrino flux as discussed next.

\section{Neutrinos from UHECRs}

Secondary neutrinos and photons can be produced by UHECRs when they interact with ambient baryonic matter and radiation fields inside the source or during their propagation from source to Earth. These particles travel in geodesics unaffected by magnetic fields and bear valuable information of the birthplace of their progenitors. The quest for sources of UHECRs has thus long been associated with the detection of neutrinos and gamma rays that might pinpoint the position of the accelerators in the sky.

The detection of these secondary particles is not straightforward however: first, the propagation of gamma rays with energy exceeding several TeV is affected by their interaction with CMB and radio photons. These interactions lead to the production of high energy electron and positron pairs which in turn up-scatter CMB or radio photons by inverse Compton processes, initiating electromagnetic cascades. As a consequence, one does not expect to observe gamma rays of energy above $\sim 100$~TeV from sources located beyond a horizon of a few Mpc. Above EeV energies, photons can again propagate over large distances, depending on the radio background, and can reach observable levels around tens of EeV. Secondary neutrinos are very useful because, unlike cosmic-rays and photons, they are not absorbed by the cosmic backgrounds while propagating through the Universe. In particular, they give a unique access to observing sources at PeV energies. However, their small interaction cross-section makes it difficult to detect them on the Earth requiring the construction of km$^3$ or larger detectors.

Neutrinos generated at UHECR sources have been investigated by a number of authors (beginning with \cite{Szabo94,Rachen98,WB99}). The normalization and the very existence of these secondaries highly depend on assumptions about the opacity of the acceleration region and on the shape of the injection spectrum as well as on the phenomenological modeling of the acceleration. For instance, \cite{WB99} obtain an estimate for the cosmic neutrino flux, by comparing the neutrino luminosity to the observed cosmic ray luminosity, in the specific case where the proton photo-meson optical depth equals unity. If the source is optically thick,  \cite{AP09} demonstrate that cosmic rays are not accelerated to the highest energies and neutrinos above $E\sim$  EeV are sharply suppressed. 

\begin{figure}[!t]
\centerline{\includegraphics[height=0.33\textheight]{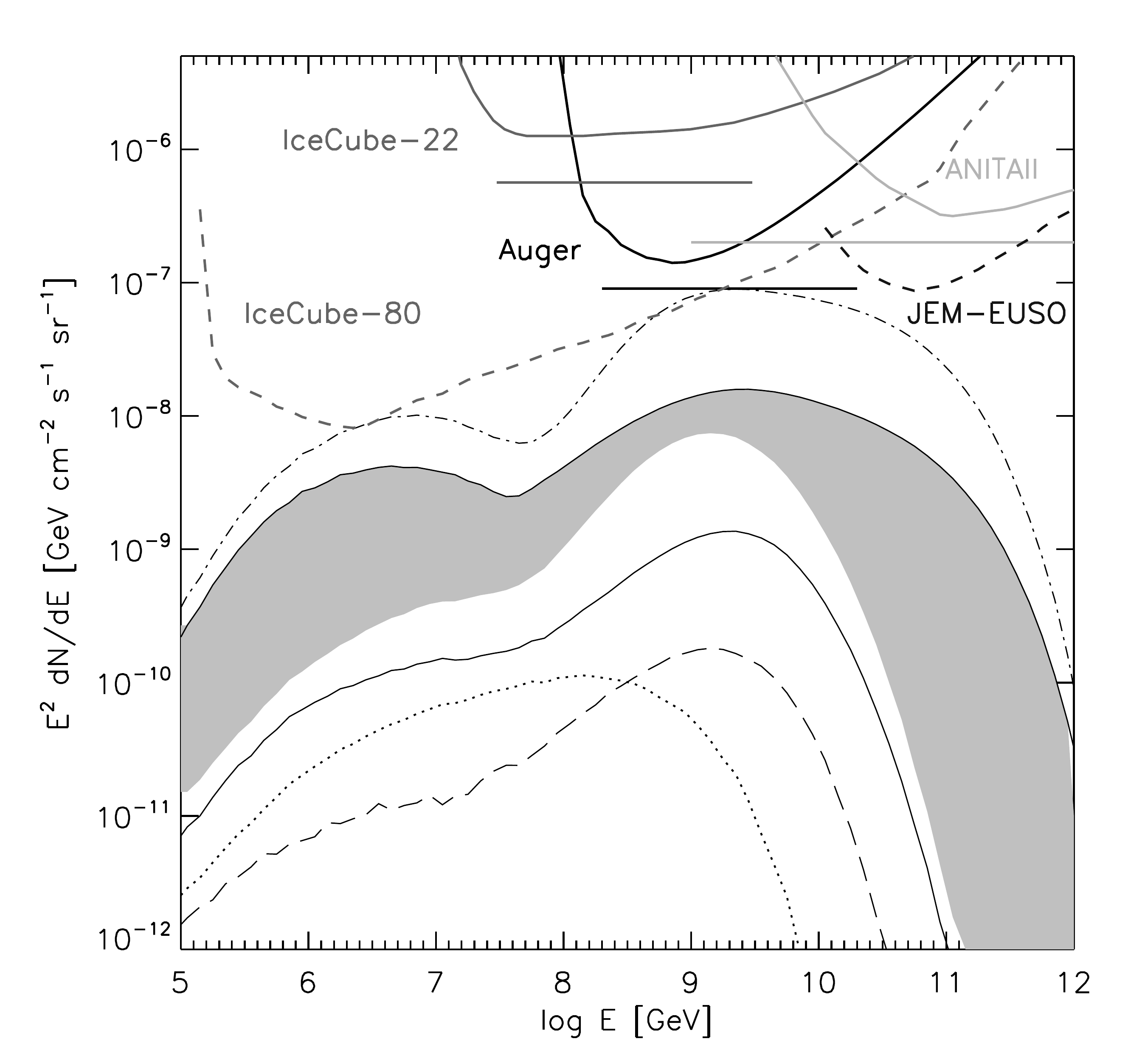}}
\caption{Cosmogenic neutrino flux for all flavors, for different UHECR parameters compared to instrument sensitivities. Dash-dot line corresponds to a strong source evolution case with a pure proton composition, dip transition model, and $E_{\rm max}=$ 3~ZeV. Uniform source evolution with iron rich (30\%) composition and $E_{Z,\rm max}<Z$ 10 EeV is shown in the dotted line and dashed line represents pure iron injection and $E_{Z,\rm max}=Z$ 100 EeV. Grey shaded range brackets dip and ankle transition models, with evolution of star formation history for $z<4$, pure proton and mixed `Galactic' compositions, and large proton $E_{\rm max} ( >  100$ EeV)). Including the uniform source evolution would broaden the shaded area down to the solid line below it. Current experimental limits (labeled lines on top) assume 90\% confidence level and full mixing neutrino oscillation. The differential limit and the integral flux limit on a pure $E^{-2}$ spectrum (straight lines) are presented for IceCube 22 lines \cite{Abbasi10}, ANITA-II \cite{ANITA10} and Auger South \cite{Auger_nu09}. For future instruments, we present the projected instrument sensitivities (dashed lines) for IceCube 80 lines (acceptances from S. Yoshida, private communication, see also \cite{Karle10}), and for JEM-EUSO \cite{JemEUSO}.}
\label{cosmoNeut}
\end{figure}

The existence of secondary neutrinos from interactions during the propagation of cosmic rays is less uncertain, but it is also subject to large variations according to the injected spectral index, chemical composition, maximum acceleration energy, and source evolution history. A number of authors have estimated the cosmogenic neutrino flux with varying assumptions (e.g., \cite{ESS01,Ave05,Seckel05,HTS05,Berezinsky06,Stanev06,Allard06,Takami09,KAO10}). Figure~\ref{cosmoNeut} summarizes the effects of different assumptions about the UHECR source evolution, the Galactic to extragalactic transition, the injected chemical composition, and $E_{\rm max}$, on the cosmogenic neutrino flux. It demonstrates that the parameter space is currently poorly constrained with uncertainties of several orders of magnitude in the predicted flux.
UHECR models with large proton $E_{\rm max} ( >  100$ EeV), source evolution corresponding to the star formation history or the GRB rate evolution,  dip or ankle transition models, and pure proton or mixed `Galactic' compositions are shaded in grey in Figure \ref{cosmoNeut} and give detectable fluxes in the EeV range with $0.06-0.2$ neutrino per year at IceCube and $0.03-0.06$ neutrino per year for the Auger Observatory. If EeV neutrinos are detected, PeV information can help select between competing models of cosmic ray composition at the highest energy and the Galactic to extragalactic transition at ankle energies.  With improved sensitivity, ZeV (=$10^{21}$ eV) neutrino observatories, such as ANITA and JEM-EUSO could explore the maximum acceleration energy. 

Due to the delay induced by EGMF on charged cosmic rays, secondary neutrinos and photons should not be detected in time coincidence with UHECRs if the sources are not continuously emitting particles, but are transient such as GRBs and young magnetars.

\section{Conclusion}

The possibility of observing high energy neutrinos is intimately related to the resolution of the long standing mystery of the origin of UHECRs. To discover the origin of UHECRs will require a coordinated approach on three complementary fronts: the direct UHECR frontier, the transition region between the knee and the ankle, and the multi-messenger interface with high-energy photons and neutrinos.

Current data suggest that watershed anisotropies will only become clear above 60 EeV and that very large statistics with good angular and energy resolution will be required. The Auger Observatory (located in Mendoza, Argentina),  will add $7\times10^3 km^2.sr$ each year of exposure to the southern sky, while the Telescope Array (located in Utah, USA) will add about $2\times10^3 km^2.sr$ each year in the North. Current technologies can reach a goal of another order of magnitude if deployed by bold scientists over very large areas (e.g., Auger North). New technologies may ease the need for large number of detector units to cover similarly large areas. Future space observatories (e.g., JEM-EUSO, OWL, Super-EUSO) promise a new avenue to reach the necessary high statistics especially if improved photon detection technologies are achieved.

Existing and upcoming high energy neutrino detectors roughly cover three energy ranges: PeV (= $10^{15}$ eV), EeV (= $10^{18}$ eV), and ZeV (= $10^{21}$ eV).  ANTARES, IceCube, and the future KM3Net are large water or ice cubic detectors that aim at observing events around PeV energies. IceCube will also have a very good sensitivity at higher energies, and will ultimately be able to cover a wide energy range from about 1~PeV to $\sim10$~EeV.  Experiments primarily dedicated to the detection of cosmic rays like the Pierre Auger Observatory and the Telescope Array have their best neutrino sensitivities in the EeV energy range.  The radio telescope ANITA and the fluorescence telescope JEM-EUSO are most effective at the highest energy neutrinos around 0.1 ZeV. 

With a significant increase in the integrated exposure to neutrinos and cosmic rays, next generation observatories may reach the sensitivity necessary to achieve charged particle astronomy and to observe ultrahigh energy photons and neutrinos, which will further illuminate the workings of the universe at the most extreme energies.


\begin{thebibliography}{10}
\expandafter\ifx\csname url\endcsname\relax
  \def\url#1{\texttt{#1}}\fi
\expandafter\ifx\csname urlprefix\endcsname\relax\def\urlprefix{URL }\fi
\expandafter\ifx\csname href\endcsname\relax
  \def\href#1#2{#2} \def\path#1{#1}\fi

\bibitem{BZ69}
V.~S. {Berezinsky}, G.~T. {Zatsepin}, {Cosmic rays at ultrahigh energies
  (neutrino?).}, Physics Letters B 28 (1969) 423--424.
\newblock \href {http://dx.doi.org/10.1016/0370-2693(69)90341-4}
  {\path{doi:10.1016/0370-2693(69)90341-4}}.

\bibitem{Stecker79}
F.~W. {Stecker}, {Diffuse fluxes of cosmic high-energy neutrinos}, ApJ 228
  (1979) 919--927.
\newblock \href {http://dx.doi.org/10.1086/156919} {\path{doi:10.1086/156919}}.

\bibitem{AM09}
L.~A. {Anchordoqui}, T.~{Montaruli}, {In Search for Extraterrestrial High
  Energy Neutrinos}, Ann. Rev. Nucl. Part. Sci. 60~(129-162).
\newblock \href {http://arxiv.org/abs/0912.1035} {\path{arXiv:0912.1035}}.

\bibitem{G66}
K.~Greisen, End to the cosmic-ray spectrum?, Phys. Rev. Lett. 16 (1966)
  748--750.
\newblock \href {http://dx.doi.org/10.1103/PhysRevLett.16.748}
  {\path{doi:10.1103/PhysRevLett.16.748}}.

\bibitem{ZK66}
G.~Zatsepin, V.~Kuzmin, Upper limit of the spectrum of cosmic rays, J. Exp.
  Theor. Phys. Lett. 4 (1966) 78--80.

\bibitem{Abbasi09}
R.~U. {Abbasi}, et~al., {Measurement of the flux of ultra high energy cosmic
  rays by the stereo technique}, Astroparticle Physics 32 (2009) 53--60.
\newblock \href {http://arxiv.org/abs/0904.4500} {\path{arXiv:0904.4500}},
  \href {http://dx.doi.org/10.1016/j.astropartphys.2009.06.001}
  {\path{doi:10.1016/j.astropartphys.2009.06.001}}.

\bibitem{Abraham:2008ru}
J.~{Abraham}, et~al., {Observation of the suppression of the flux of cosmic
  rays above $4\times 10^{19}$eV}, Phys. Rev. Lett. 101 (2008) 061101.
\newblock \href {http://arxiv.org/abs/0806.4302} {\path{arXiv:0806.4302}},
  \href {http://dx.doi.org/10.1103/PhysRevLett.101.061101}
  {\path{doi:10.1103/PhysRevLett.101.061101}}.

\bibitem{Auger1}
J.~{Abraham}, et~al., {Correlation of the Highest-Energy Cosmic Rays with
  Nearby Extragalactic Objects}, Science 318 (2007) 938.
\newblock \href {http://arxiv.org/abs/0711.2256} {\path{arXiv:0711.2256}},
  \href {http://dx.doi.org/10.1126/science.1151124}
  {\path{doi:10.1126/science.1151124}}.

\bibitem{Auger2}
J.~{Abraham}, et~al., {Correlation of the highest-energy cosmic rays with the
  positions of nearby active galactic nuclei}, Astroparticle Physics 29 (2008)
  188--204.
\newblock \href {http://dx.doi.org/10.1016/j.astropartphys.2008.01.002}
  {\path{doi:10.1016/j.astropartphys.2008.01.002}}.

\bibitem{Abraham:2010yv}
J.~{Abraham}, et~al., {Measurement of the Depth of Maximum of Extensive Air
  Showers above EeV}, Phys. Rev. Lett. 104 (2010) 091101.
\newblock \href {http://arxiv.org/abs/1002.0699} {\path{arXiv:1002.0699}},
  \href {http://dx.doi.org/10.1103/PhysRevLett.104.091101}
  {\path{doi:10.1103/PhysRevLett.104.091101}}.

\bibitem{KKAO11}
K.~Kotera, A.~V. Olinto, \href{http://arxiv.org/abs/1101.4256v1}{The
  astrophysics of ultrahigh energy cosmic rays}\href
  {http://arxiv.org/abs/1101.4256v1} {\path{arXiv:1101.4256v1}}.
\newline\urlprefix\url{http://arxiv.org/abs/1101.4256v1}

\bibitem{Letessier11}
A.~{Letessier-Selvon}, T.~{Stanev}, Ultrahigh energy cosmic rays, Rev. Mod.
  Phys.

\bibitem{Takeda98}
M.~{Takeda}, et~al., {Extension of the Cosmic-Ray Energy Spectrum beyond the
  Predicted Greisen-Zatsepin-Kuz'min Cutoff}, Physical Review Letters 81 (1998)
  1163--1166.
\newblock \href {http://arxiv.org/abs/arXiv:astro-ph/9807193}
  {\path{arXiv:arXiv:astro-ph/9807193}}, \href
  {http://dx.doi.org/10.1103/PhysRevLett.81.1163}
  {\path{doi:10.1103/PhysRevLett.81.1163}}.

\bibitem{Abraham10}
J.~{Abraham}, et~al., {Measurement of the energy spectrum of cosmic rays above
  10^{18} eV using the Pierre Auger Observatory}, Physics Letters B 685 (2010)
  239--246.
\newblock \href {http://arxiv.org/abs/1002.1975} {\path{arXiv:1002.1975}},
  \href {http://dx.doi.org/10.1016/j.physletb.2010.02.013}
  {\path{doi:10.1016/j.physletb.2010.02.013}}.

\bibitem{Allard07}
D.~{Allard}, E.~{Parizot}, A.~V. {Olinto}, {On the transition from galactic to
  extragalactic cosmic-rays: Spectral and composition features from two
  opposite scenarios}, Astroparticle Physics 27 (2007) 61--75.
\newblock \href {http://dx.doi.org/10.1016/j.astropartphys.2006.09.006}
  {\path{doi:10.1016/j.astropartphys.2006.09.006}}.

\bibitem{BG88}
V.~S. {Berezinsky}, S.~I. {Grigorieva}, {A bump in the ultra-high energy cosmic
  ray spectrum}, A\&A 199 (1988) 1--2.

\bibitem{BGG06}
V.~{Berezinsky}, A.~{Gazizov}, S.~{Grigorieva}, {On astrophysical solution to
  ultrahigh energy cosmic rays}, Phys. Rev. D 74~(4) (2006) 043005--+.
\newblock \href {http://arxiv.org/abs/arXiv:hep-ph/0204357}
  {\path{arXiv:arXiv:hep-ph/0204357}}, \href
  {http://dx.doi.org/10.1103/PhysRevD.74.043005}
  {\path{doi:10.1103/PhysRevD.74.043005}}.

\bibitem{VC06}
M.-P. {V{\'e}ron-Cetty}, P.~{V{\'e}ron}, {A catalogue of quasars and active
  nuclei: 12th edition}, A\&A 455 (2006) 773--777.
\newblock \href {http://dx.doi.org/10.1051/0004-6361:20065177}
  {\path{doi:10.1051/0004-6361:20065177}}.

\bibitem{Abreu10}
P.~{Abreu}, et~al., {Update on the correlation of the highest energy cosmic
  rays with nearby extragalactic matter}, Astroparticle Physics 34 (2010) 314.
\newblock \href {http://arxiv.org/abs/1009.1855} {\path{arXiv:1009.1855}}.

\bibitem{Abbasi08corr}
R.~U. {Abbasi}, et~al., {Search for correlations between HiRes stereo events
  and active galactic nuclei}, Astroparticle Physics 30 (2008) 175--179.
\newblock \href {http://arxiv.org/abs/0804.0382} {\path{arXiv:0804.0382}},
  \href {http://dx.doi.org/10.1016/j.astropartphys.2008.08.004}
  {\path{doi:10.1016/j.astropartphys.2008.08.004}}.

\bibitem{Aglietta:2007}
J.~Abraham, et~al., {Upper limit on the cosmic-ray photon flux above 10$^{19}$
  eV using the surface detector of the Pierre Auger Observatory}, Astropart.
  Phys. 29 (2008) 243--256.
\newblock \href {http://arxiv.org/abs/0712.1147} {\path{arXiv:0712.1147}},
  \href {http://dx.doi.org/10.1016/j.astropartphys.2008.01.003}
  {\path{doi:10.1016/j.astropartphys.2008.01.003}}.

\bibitem{Abraham:2009qb}
J.~{Abraham}, et~al., {Upper limit on the cosmic-ray photon fraction at EeV
  energies from the Pierre Auger Observatory}, Astropart. Phys. 31 (2009)
  399--406.
\newblock \href {http://arxiv.org/abs/0903.1127} {\path{arXiv:0903.1127}},
  \href {http://dx.doi.org/10.1016/j.astropartphys.2009.04.003}
  {\path{doi:10.1016/j.astropartphys.2009.04.003}}.

\bibitem{Auger_nu09}
J.~{Abraham}, et~al., {Limit on the diffuse flux of ultrahigh energy tau
  neutrinos with the surface detector of the Pierre Auger Observatory},
  Phys.~Rev.~D 79~(10) (2009) 102001.
\newblock \href {http://arxiv.org/abs/0903.3385} {\path{arXiv:0903.3385}},
  \href {http://dx.doi.org/10.1103/PhysRevD.79.102001}
  {\path{doi:10.1103/PhysRevD.79.102001}}.

\bibitem{Abbasi08neu}
R.~U. {Abbasi}, et~al., {An Upper Limit on the Electron-Neutrino Flux from the
  HiRes Detector}, Astrophys. J. 684 (2008) 790--793.
\newblock \href {http://arxiv.org/abs/0803.0554} {\path{arXiv:0803.0554}},
  \href {http://dx.doi.org/10.1086/590335} {\path{doi:10.1086/590335}}.

\bibitem{Szabo94}
A.~P. {Szabo}, R.~J. {Protheroe}, {Implications of particle acceleration in
  active galactic nuclei for cosmic rays and high energy neutrino astronomy},
  Astroparticle Physics 2 (1994) 375--392.
\newblock \href {http://arxiv.org/abs/arXiv:astro-ph/9405020}
  {\path{arXiv:arXiv:astro-ph/9405020}}, \href
  {http://dx.doi.org/10.1016/0927-6505(94)90027-2}
  {\path{doi:10.1016/0927-6505(94)90027-2}}.

\bibitem{Rachen98}
J.~P. {Rachen}, P.~{M{\'e}sz{\'a}ros}, {Photohadronic neutrinos from transients
  in astrophysical sources}, Phys.~Rev.~D 58~(12) (1998) 123005--+.
\newblock \href {http://arxiv.org/abs/arXiv:astro-ph/9802280}
  {\path{arXiv:arXiv:astro-ph/9802280}}, \href
  {http://dx.doi.org/10.1103/PhysRevD.58.123005}
  {\path{doi:10.1103/PhysRevD.58.123005}}.

\bibitem{WB99}
E.~{Waxman}, J.~{Bahcall}, {High energy neutrinos from astrophysical sources:
  An upper bound}, Phys. Rev. D 59~(2) (1999) 023002--+.
\newblock \href {http://arxiv.org/abs/arXiv:hep-ph/9807282}
  {\path{arXiv:arXiv:hep-ph/9807282}}, \href
  {http://dx.doi.org/10.1103/PhysRevD.59.023002}
  {\path{doi:10.1103/PhysRevD.59.023002}}.

\bibitem{Mucke99}
A.~{M\"ucke}, et~al., {Photohadronic processes in astrophysical environments},
  Publications of the Astronomical Society of Australia 16 (1999) 160--6.
\newblock \href {http://arxiv.org/abs/arXiv:astro-ph/9808279}
  {\path{arXiv:arXiv:astro-ph/9808279}}.

\bibitem{M00}
A.~{M{\"u}cke}, et~al., {Monte Carlo simulations of photohadronic processes in
  astrophysics}, Computer Physics Communications 124 (2000) 290--314.
\newblock \href {http://arxiv.org/abs/arXiv:astro-ph/9903478}
  {\path{arXiv:arXiv:astro-ph/9903478}}, \href
  {http://dx.doi.org/10.1016/S0010-4655(99)00446-4}
  {\path{doi:10.1016/S0010-4655(99)00446-4}}.

\bibitem{Anchordoqui08}
L.~A. {Anchordoqui}, D.~{Hooper}, S.~{Sarkar}, A.~M. {Taylor}, {High energy
  neutrinos from astrophysical accelerators of cosmic ray nuclei},
  Astroparticle Physics 29 (2008) 1--13.
\newblock \href {http://arxiv.org/abs/arXiv:astro-ph/0703001}
  {\path{arXiv:arXiv:astro-ph/0703001}}, \href
  {http://dx.doi.org/10.1016/j.astropartphys.2007.10.006}
  {\path{doi:10.1016/j.astropartphys.2007.10.006}}.

\bibitem{Kachelriess08}
M.~{Kachelrie{\ss}}, S.~{Ostapchenko}, R.~{Tom{\`a}s}, {High energy neutrino
  yields from astrophysical sources. II. Magnetized sources}, Phys.~Rev.~D
  77~(2) (2008) 023007--+.
\newblock \href {http://arxiv.org/abs/0708.3047} {\path{arXiv:0708.3047}},
  \href {http://dx.doi.org/10.1103/PhysRevD.77.023007}
  {\path{doi:10.1103/PhysRevD.77.023007}}.

\bibitem{Ahlers09}
M.~{Ahlers}, L.~A. {Anchordoqui}, S.~{Sarkar}, {Neutrino diagnostics of
  ultrahigh energy cosmic ray protons}, Phys. Rev. D 79~(8) (2009) 083009--+.
\newblock \href {http://arxiv.org/abs/0902.3993} {\path{arXiv:0902.3993}},
  \href {http://dx.doi.org/10.1103/PhysRevD.79.083009}
  {\path{doi:10.1103/PhysRevD.79.083009}}.

\bibitem{AP09}
D.~{Allard}, R.~J. {Protheroe}, {Interactions of UHE cosmic ray nuclei with
  radiation during acceleration: consequences on the spectrum and composition},
  A\&A 502~(3) (2009) 803--815.
\newblock \href {http://arxiv.org/abs/0902.4538} {\path{arXiv:0902.4538}}.

\bibitem{MPR01}
K.~{Mannheim}, R.~J. {Protheroe}, J.~P. {Rachen}, {Cosmic ray bound for models
  of extragalactic neutrino production}, Phys.~Rev.~D 63~(2) (2001) 023003--+.
\newblock \href {http://arxiv.org/abs/arXiv:astro-ph/9812398}
  {\path{arXiv:arXiv:astro-ph/9812398}}, \href
  {http://dx.doi.org/10.1103/PhysRevD.63.023003}
  {\path{doi:10.1103/PhysRevD.63.023003}}.

\bibitem{Inoue07}
S.~{Inoue}, G.~{Sigl}, F.~{Miniati}, E.~{Armengaud}, {Ultrahigh energy cosmic
  rays as heavy nuclei from cluster accretion shocks}, Proceedings of the 30th
  International Cosmic Ray Conference, Merida, Mexico\href
  {http://arxiv.org/abs/0711.1027} {\path{arXiv:0711.1027}}.

\bibitem{MIN08}
K.~{Murase}, S.~{Inoue}, S.~{Nagataki}, {Cosmic Rays above the Second Knee from
  Clusters of Galaxies and Associated High-Energy Neutrino Emission}, ApJl 689
  (2008) L105--L108.
\newblock \href {http://arxiv.org/abs/0805.0104} {\path{arXiv:0805.0104}},
  \href {http://dx.doi.org/10.1086/595882} {\path{doi:10.1086/595882}}.

\bibitem{WB00}
E.~{Waxman}, J.~N. {Bahcall}, {Neutrino Afterglow from Gamma-Ray Bursts:
  \~{}10^{18} EV}, ApJ 541 (2000) 707--711.
\newblock \href {http://arxiv.org/abs/arXiv:hep-ph/9909286}
  {\path{arXiv:arXiv:hep-ph/9909286}}, \href {http://dx.doi.org/10.1086/309462}
  {\path{doi:10.1086/309462}}.

\bibitem{Dai01}
Z.~G. {Dai}, T.~{Lu}, {Neutrino Afterglows and Progenitors of Gamma-Ray
  Bursts}, ApJ 551 (2001) 249--253.
\newblock \href {http://arxiv.org/abs/arXiv:astro-ph/0002430}
  {\path{arXiv:arXiv:astro-ph/0002430}}, \href
  {http://dx.doi.org/10.1086/320056} {\path{doi:10.1086/320056}}.

\bibitem{Dermer02}
C.~D. {Dermer}, {Neutrino, Neutron, and Cosmic-Ray Production in the External
  Shock Model of Gamma-Ray Bursts}, ApJ 574 (2002) 65--87.
\newblock \href {http://arxiv.org/abs/arXiv:astro-ph/0005440}
  {\path{arXiv:arXiv:astro-ph/0005440}}, \href
  {http://dx.doi.org/10.1086/340893} {\path{doi:10.1086/340893}}.

\bibitem{Murase06}
K.~{Murase}, K.~{Ioka}, S.~{Nagataki}, T.~{Nakamura}, {High-Energy Neutrinos
  and Cosmic Rays from Low-Luminosity Gamma-Ray Bursts?}, ApJl 651 (2006)
  L5--L8.
\newblock \href {http://arxiv.org/abs/arXiv:astro-ph/0607104}
  {\path{arXiv:arXiv:astro-ph/0607104}}, \href
  {http://dx.doi.org/10.1086/509323} {\path{doi:10.1086/509323}}.

\bibitem{Murase08}
K.~{Murase}, K.~{Ioka}, S.~{Nagataki}, T.~{Nakamura}, {High-energy cosmic-ray
  nuclei from high- and low-luminosity gamma-ray bursts and implications for
  multimessenger astronomy}, Phys. Rev. D 78~(2) (2008) 023005--+.
\newblock \href {http://arxiv.org/abs/0801.2861} {\path{arXiv:0801.2861}},
  \href {http://dx.doi.org/10.1103/PhysRevD.78.023005}
  {\path{doi:10.1103/PhysRevD.78.023005}}.

\bibitem{MI08}
K.~{Murase}, K.~{Ioka}, {Closure Relations for e^{+/-} Pair Signatures in
  Gamma-Ray Bursts}, ApJ 676 (2008) 1123--1129.
\newblock \href {http://arxiv.org/abs/0708.1370} {\path{arXiv:0708.1370}},
  \href {http://dx.doi.org/10.1086/527667} {\path{doi:10.1086/527667}}.

\bibitem{Murase09}
K.~{Murase}, P.~{M{\'e}sz{\'a}ros}, B.~{Zhang}, {Probing the birth of fast
  rotating magnetars through high-energy neutrinos}, Phys.~Rev.~D 79~(10)
  (2009) 103001--+.
\newblock \href {http://arxiv.org/abs/0904.2509} {\path{arXiv:0904.2509}},
  \href {http://dx.doi.org/10.1103/PhysRevD.79.103001}
  {\path{doi:10.1103/PhysRevD.79.103001}}.

\bibitem{BBP97}
V.~S. {Berezinsky}, P.~{Blasi}, V.~S. {Ptuskin}, {Clusters of Galaxies as
  Storage Room for Cosmic Rays}, ApJ 487 (1997) 529--+.
\newblock \href {http://arxiv.org/abs/arXiv:astro-ph/9609048}
  {\path{arXiv:arXiv:astro-ph/9609048}}, \href
  {http://dx.doi.org/10.1086/304622} {\path{doi:10.1086/304622}}.

\bibitem{CB98}
S.~{Colafrancesco}, P.~{Blasi}, {Clusters of galaxies and the diffuse gamma-ray
  background}, Astroparticle Physics 9 (1998) 227--246.
\newblock \href {http://arxiv.org/abs/arXiv:astro-ph/9804262}
  {\path{arXiv:arXiv:astro-ph/9804262}}, \href
  {http://dx.doi.org/10.1016/S0927-6505(98)00018-8}
  {\path{doi:10.1016/S0927-6505(98)00018-8}}.

\bibitem{RGD04}
C.~{Rordorf}, D.~{Grasso}, K.~{Dolag}, {Diffusion of ultra-high energy protons
  in galaxy clusters and secondary X- and gamma-ray emissions}, Astroparticle
  Physics 22 (2004) 167--181.
\newblock \href {http://arxiv.org/abs/arXiv:astro-ph/0405046}
  {\path{arXiv:arXiv:astro-ph/0405046}}, \href
  {http://dx.doi.org/10.1016/j.astropartphys.2004.07.001}
  {\path{doi:10.1016/j.astropartphys.2004.07.001}}.

\bibitem{demarco06}
D.~{de Marco}, P.~{Hansen}, T.~{Stanev}, P.~{Blasi}, {High energy neutrinos
  from cosmic ray interactions in clusters of galaxies}, Phys. Rev. D 73~(4)
  (2006) 043004--+.
\newblock \href {http://arxiv.org/abs/arXiv:astro-ph/0511535}
  {\path{arXiv:arXiv:astro-ph/0511535}}, \href
  {http://dx.doi.org/10.1103/PhysRevD.73.043004}
  {\path{doi:10.1103/PhysRevD.73.043004}}.

\bibitem{ASM06}
E.~{Armengaud}, G.~{Sigl}, F.~{Miniati}, {Secondary gamma rays from ultrahigh
  energy cosmic rays produced in magnetized environments}, Phys. Rev. D 73~(8)
  (2006) 083008--+.
\newblock \href {http://dx.doi.org/10.1103/PhysRevD.73.083008}
  {\path{doi:10.1103/PhysRevD.73.083008}}.

\bibitem{Wolfe08}
B.~{Wolfe}, F.~{Melia}, R.~M. {Crocker}, R.~R. {Volkas}, {Neutrinos and Gamma
  Rays from Galaxy Clusters}, ApJ 687 (2008) 193--201.
\newblock \href {http://arxiv.org/abs/0807.0794} {\path{arXiv:0807.0794}},
  \href {http://dx.doi.org/10.1086/591723} {\path{doi:10.1086/591723}}.

\bibitem{KAM09}
K.~{Kotera}, D.~{Allard}, K.~{Murase}, J.~{Aoi}, Y.~{Dubois}, T.~{Pierog},
  S.~{Nagataki}, {Propagation of Ultrahigh Energy Nuclei in Clusters of
  Galaxies: Resulting Composition and Secondary Emissions}, ApJ 707 (2009)
  370--386.
\newblock \href {http://arxiv.org/abs/0907.2433} {\path{arXiv:0907.2433}},
  \href {http://dx.doi.org/10.1088/0004-637X/707/1/370}
  {\path{doi:10.1088/0004-637X/707/1/370}}.

\bibitem{ESS01}
R.~{Engel}, D.~{Seckel}, T.~{Stanev}, {Neutrinos from propagation of ultrahigh
  energy protons}, Phys. Rev. D 64~(9) (2001) 093010--+.
\newblock \href {http://arxiv.org/abs/arXiv:astro-ph/0101216}
  {\path{arXiv:arXiv:astro-ph/0101216}}, \href
  {http://dx.doi.org/10.1103/PhysRevD.64.093010}
  {\path{doi:10.1103/PhysRevD.64.093010}}.

\bibitem{Ave05}
M.~{Ave}, et~al., {Cosmogenic neutrinos from ultra-high energy nuclei},
  Astropart. Phys. 23 (2005) 19--29.
\newblock \href {http://arxiv.org/abs/arXiv:astro-ph/0409316}
  {\path{arXiv:arXiv:astro-ph/0409316}}, \href
  {http://dx.doi.org/10.1016/j.astropartphys.2004.11.001}
  {\path{doi:10.1016/j.astropartphys.2004.11.001}}.

\bibitem{Seckel05}
D.~{Seckel}, T.~{Stanev}, {Neutrinos: The Key to Ultrahigh Energy Cosmic Rays},
  Physical Review Letters 95~(14) (2005) 141101--+.
\newblock \href {http://arxiv.org/abs/arXiv:astro-ph/0502244}
  {\path{arXiv:arXiv:astro-ph/0502244}}, \href
  {http://dx.doi.org/10.1103/PhysRevLett.95.141101}
  {\path{doi:10.1103/PhysRevLett.95.141101}}.

\bibitem{HTS05}
D.~{Hooper}, A.~{Taylor}, S.~{Sarkar}, {The impact of heavy nuclei on the
  cosmogenic neutrino flux}, Astroparticle Physics 23 (2005) 11--17.
\newblock \href {http://arxiv.org/abs/arXiv:astro-ph/0407618}
  {\path{arXiv:arXiv:astro-ph/0407618}}, \href
  {http://dx.doi.org/10.1016/j.astropartphys.2004.11.002}
  {\path{doi:10.1016/j.astropartphys.2004.11.002}}.

\bibitem{Berezinsky06}
V.~{Berezinsky}, {Ultra High Energy Neutrino Astronomy}, Nuclear Physics B
  Proceedings Supplements 151 (2006) 260--269.
\newblock \href {http://arxiv.org/abs/arXiv:astro-ph/0505220}
  {\path{arXiv:arXiv:astro-ph/0505220}}, \href
  {http://dx.doi.org/10.1016/j.nuclphysbps.2005.07.062}
  {\path{doi:10.1016/j.nuclphysbps.2005.07.062}}.

\bibitem{Stanev06}
T.~{Stanev}, D.~{de Marco}, M.~A. {Malkan}, F.~W. {Stecker}, {Cosmogenic
  neutrinos from cosmic ray interactions with extragalactic infrared photons},
  Phys.~Rev.~D 73~(4) (2006) 043003--+.
\newblock \href {http://dx.doi.org/10.1103/PhysRevD.73.043003}
  {\path{doi:10.1103/PhysRevD.73.043003}}.

\bibitem{Allard06}
D.~{Allard}, M.~{Ave}, N.~{Busca}, M.~A. {Malkan}, A.~V. {Olinto},
  E.~{Parizot}, F.~W. {Stecker}, T.~{Yamamoto}, {Cosmogenic neutrinos from the
  propagation of ultrahigh energy nuclei}, JCAP 9 (2006) 5.
\newblock \href {http://arxiv.org/abs/arXiv:astro-ph/0605327}
  {\path{arXiv:arXiv:astro-ph/0605327}}, \href
  {http://dx.doi.org/10.1088/1475-7516/2006/09/005}
  {\path{doi:10.1088/1475-7516/2006/09/005}}.

\bibitem{Takami09}
H.~{Takami}, K.~{Murase}, S.~{Nagataki}, K.~{Sato}, {Cosmogenic neutrinos as a
  probe of the transition from Galactic to extragalactic cosmic rays},
  Astroparticle Physics 31 (2009) 201--211.
\newblock \href {http://dx.doi.org/10.1016/j.astropartphys.2009.01.006}
  {\path{doi:10.1016/j.astropartphys.2009.01.006}}.

\bibitem{KAO10}
K.~{Kotera}, D.~{Allard}, A.~V. {Olinto}, {Cosmogenic neutrinos: parameter
  space and detectabilty from PeV to ZeV}, JCAP 10 (2010) 13.
\newblock \href {http://arxiv.org/abs/1009.1382} {\path{arXiv:1009.1382}},
  \href {http://dx.doi.org/10.1088/1475-7516/2010/10/013}
  {\path{doi:10.1088/1475-7516/2010/10/013}}.

\bibitem{Wall05}
J.~V. {Wall}, C.~A. {Jackson}, P.~A. {Shaver}, I.~M. {Hook}, K.~I.
  {Kellermann}, {The Parkes quarter-Jansky flat-spectrum sample. III. Space
  density and evolution of QSOs}, A\&A 434 (2005) 133--148.
\newblock \href {http://arxiv.org/abs/arXiv:astro-ph/0408122}
  {\path{arXiv:arXiv:astro-ph/0408122}}, \href
  {http://dx.doi.org/10.1051/0004-6361:20041786}
  {\path{doi:10.1051/0004-6361:20041786}}.

\bibitem{Abbasi10}
R.~U. {Abbasi}, et~al., {Indications of Proton-Dominated Cosmic-Ray Composition
  above 1.6 EeV}, Physical Review Letters 104~(16) (2010) 161101.
\newblock \href {http://arxiv.org/abs/0910.4184} {\path{arXiv:0910.4184}},
  \href {http://dx.doi.org/10.1103/PhysRevLett.104.161101}
  {\path{doi:10.1103/PhysRevLett.104.161101}}.

\bibitem{ANITA10}
P.~W. {Gorham}, et~al., {Observational Constraints on the Ultra-high Energy
  Cosmic Neutrino Flux from the Second Flight of the ANITA Experiment},
  arXiv:1003.2961\href {http://arxiv.org/abs/1003.2961}
  {\path{arXiv:1003.2961}}.

\bibitem{Karle10}
A.~{Karle}, {IceCube}, arXiv: 1003.5715\href {http://arxiv.org/abs/1003.5715}
  {\path{arXiv:1003.5715}}.

\bibitem{JemEUSO}
Y.~{Takahashi}, et~al., {The Jem-Euso Mission}, New J. Phys. 11 (2009) 065009.

\end{thebibliography}


\begin{thebibliography}{9}

\bibitem{BZ69}
V.~S. {Berezinsky}, G.~T. {Zatsepin}, Physics Letters B 28 (1969) 423.

\bibitem{Stecker79}
F.~W. {Stecker},  ApJ 228 (1979) 919.

\bibitem{AM09}
L.~A. {Anchordoqui}, T.~{Montaruli}, Ann. Rev. Nucl. Part. Sci. 60, 129.

\bibitem{G66}
K.~Greisen, Phys. Rev. Lett. 16 (1966) 748.

\bibitem{ZK66}
G.~Zatsepin, V.~Kuzmin,  J. Exp.  Theor. Phys. Lett. 4 (1966) 78.

\bibitem{Abbasi09}
R.~U. {Abbasi}, et~al., Astroparticle Physics 32 (2009) 53.

\bibitem{Abraham:2008ru}
J.~{Abraham}, et~al., Phys. Rev. Lett. 101 (2008) 061101.

\bibitem{Auger1}
J.~{Abraham}, et~al.,  Science 318 (2007) 938.

\bibitem{Auger2}
J.~{Abraham}, et~al.,  Astroparticle Physics 29 (2008)  188.

\bibitem{Abraham:2010yv}
J.~{Abraham}, et~al., Phys. Rev. Lett. 104 (2010) 091101.

\bibitem{KKAO11}
K.~Kotera, A.~V. Olinto, Annu. Rev. Astron. Astrophys. v. 49 (2011), arXiv:1101.4256v1

\bibitem{Letessier11}
A.~{Letessier-Selvon}, T.~{Stanev}, Rev. Mod. Phys. (2011).

\bibitem{Takeda98}
M.~{Takeda}, et~al., Physical Review Letters 81 (1998)  1163.

\bibitem{Abraham10}
J.~{Abraham}, et~al., Physics Letters B 685 (2010) 239.

\bibitem{Allard07}
D.~{Allard}, E.~{Parizot}, A.~V. {Olinto}, Astroparticle Physics 27 (2007) 61.

\bibitem{BG88}
V.~S. {Berezinsky}, S.~I. {Grigorieva},  A\&A 199 (1988) 1.

\bibitem{BGG06}
V.~{Berezinsky}, A.~{Gazizov}, S.~{Grigorieva}, Phys. Rev. D 74~(4) (2006) 043005

\bibitem{VC06}
M.-P. {V{\'e}ron-Cetty}, P.~{V{\'e}ron},  A\&A 455 (2006) 773.

\bibitem{Abreu10}
P.~{Abreu}, et~al., Astroparticle Physics 34 (2010) 314.

\bibitem{Abbasi08corr}
R.~U. {Abbasi}, et~al., Astroparticle Physics 30 (2008) 175.

\bibitem{Aglietta:2007}
J.~Abraham, et~al.,  Astropart.  Phys. 29 (2008) 243.

\bibitem{Abraham:2009qb}
J.~{Abraham}, et~al., Astropart. Phys. 31 (2009)  399.

\bibitem{Auger_nu09}
J.~{Abraham}, et~al., Phys.~Rev.~D 79~(10) (2009) 102001.

\bibitem{Abbasi08neu}
R.~U. {Abbasi}, et~al., Astrophys. J. 684 (2008) 790.

\bibitem{Szabo94}
A.~P. {Szabo}, R.~J. {Protheroe},   Astroparticle Physics 2 (1994) 375.

\bibitem{Rachen98}
J.~P. {Rachen}, P.~{M{\'e}sz{\'a}ros}, Phys.~Rev.~D 58~(12) (1998) 123005

\bibitem{WB99}
E.~{Waxman}, J.~{Bahcall},  Phys. Rev. D 59~(2) (1999) 023002


\bibitem{AP09}
D.~{Allard}, R.~J. {Protheroe},  A\&A 502~(3) (2009) 803.

\bibitem{ESS01}
R.~{Engel}, D.~{Seckel}, T.~{Stanev},  Phys. Rev. D 64~(9) (2001) 093010

\bibitem{Ave05}
M.~{Ave}, et~al.,   Astropart. Phys. 23 (2005) 19.

\bibitem{Seckel05}
D.~{Seckel}, T.~{Stanev}, Physical Review Letters 95~(14) (2005) 141101
  
\bibitem{HTS05}
D.~{Hooper}, A.~{Taylor}, S.~{Sarkar}, Astroparticle Physics 23 (2005) 11.

\bibitem{Berezinsky06}
V.~{Berezinsky}, Nuclear Physics B
  Proceedings Supplements 151 (2006) 260.

\bibitem{Stanev06}
T.~{Stanev}, D.~{de Marco}, M.~A. {Malkan}, F.~W. {Stecker}, 
  Phys.~Rev.~D 73~(4) (2006) 043003.

\bibitem{Allard06}
D.~{Allard}, M.~{Ave}, N.~{Busca}, M.~A. {Malkan}, A.~V. {Olinto},
  E.~{Parizot}, F.~W. {Stecker}, T.~{Yamamoto}, JCAP 9 (2006) 5.

\bibitem{Takami09}
H.~{Takami}, K.~{Murase}, S.~{Nagataki}, K.~{Sato},
  Astroparticle Physics 31 (2009) 201.

\bibitem{KAO10}
K.~{Kotera}, D.~{Allard}, A.~V. {Olinto}, JCAP 10 (2010) 13.


\bibitem{Abbasi10}
R.~U. {Abbasi}, et~al., Physical Review Letters 104~(16) (2010) 161101.

\bibitem{ANITA10}
P.~W. {Gorham}, et~al.,  arXiv:1003.2961

\bibitem{Karle10}
A.~{Karle}, {IceCube}, arXiv: 1003.5715

\bibitem{JemEUSO}
Y.~{Takahashi}, et~al., {The Jem-Euso Mission}, New J. Phys. 11 (2009) 065009.

\end{thebibliography}
\end{document}